\documentclass[twocolumn,twocolappendix]{aastex631}

\usepackage{xspace}
\usepackage{hyperref}
\usepackage{multirow}
\usepackage{graphicx}
\usepackage{enumitem}
\usepackage{amssymb}
\usepackage{amsmath}
\usepackage{booktabs}
\usepackage{pifont}

\newcommand{\xmm}{XMM-\textit{Newton}\xspace}
\newcommand{\nustar}{\textit{NuSTAR}\xspace}

\newcommand{\todo}{\ifmmode \text{\color{red}\Huge{\(\bullet\)}} \else {\color{red}{\Huge$\bullet$}}\fi}

\def\fexxvi{Fe\,{\sc xxvi}}
\def\fexxv{Fe\,{\sc xxv}}

\begin{document}

\title{NuSTAR Detection of an Absorption Feature in ESP~39607: Evidence for an Ultra-Fast Inflow?}

\correspondingauthor{Alessandro Peca}
\email{$^{\star}$peca.alessandro@gmail.com}

\author[0000-0003-2196-3298]{Alessandro Peca$^{\star}$}
\affiliation{Eureka Scientific, 2452 Delmer Street, Suite 100, Oakland, CA 94602-3017, USA}
\affiliation{Department of Physics, Yale University, P.O. Box 208120, New Haven, CT 06520, USA}

\author[0000-0002-7998-9581]{Michael J. Koss}
\affiliation{Eureka Scientific, 2452 Delmer Street, Suite 100, Oakland, CA 94602-3017, USA}
\affiliation{Space Science Institute, 4750 Walnut Street, Suite 205, Boulder, CO 80301, USA}

\author[0000-0003-1200-5071]{Roberto Serafinelli}
\affiliation{Instituto de Estudios Astrof\'{\i}sicos, Facultad de Ingenier\'{\i}a y Ciencias, Universidad Diego Portales, Av. Ej\'{e}cito Libertador 441,
Santiago, Chile}
\affiliation{INAF - Osservatorio Astronomico di Roma, Via Frascati 33, 00078, Monte Porzio Catone (Roma), Italy}

\author[0000-0001-5231-2645]{Claudio Ricci}
\affiliation{Instituto de Estudios Astrof\'{\i}sicos, Facultad de Ingenier\'{\i}a y Ciencias, Universidad Diego Portales, Av. Ej\'{e}cito Libertador 441,
Santiago, Chile}

\author[0000-0002-0745-9792]{C. Megan Urry}
\affiliation{Department of Physics, Yale University, P.O. Box 208120, New Haven, CT 06520, USA}
\affiliation{Yale Center for Astronomy \& Astrophysics, 52 Hillhouse Avenue, New Haven, CT 06511, USA}

\author[0000-0002-5273-4634]{Giulia Cerini}
\affiliation{Jet Propulsion Laboratory, California Institute of Technology, 4800 Oak Grove Dr, Pasadena, CA 91109, USA}

\author[0000-0001-9379-4716]{Peter G. Boorman}
\affiliation{Cahill Center for Astronomy and Astrophysics, California Institute of Technology, Pasadena, CA 91125, USA}



\begin{abstract}

We report the serendipitous discovery of an absorption feature at 4.8 keV in the \nustar\ spectra of ESP~39607, a Seyfert 2 galaxy at $z = 0.201$, observed in May 2023 and August 2024. The feature is detected in both observations with individual significance levels between 2 and 3$\sigma$, computed with multiple statistical methods. The combined probability of detecting it in both observations is $\gtrsim$4$\sigma$. 
The absorption feature is consistent with an ultra-fast inflow (UFI) potentially associated with \fexxv\ or \fexxvi\ K$\alpha$ transitions. The inferred inflow velocity is $\sim$0.15--0.20$c$, with an estimated launching radius of 22--89 $R_g$, depending on the assumed iron transition and whether radiation pressure is accounted for. Photoionization modeling associates the UFI primarily with \fexxv\ K$\alpha$ absorption, blended with a minor contribution from \fexxvi\ K$\alpha$.
Alternative explanations, including associations with the warm-hot intergalactic medium or outflows of lighter elements, are investigated but found unlikely. If confirmed, this detection represents a rare example of a UFI, providing valuable evidence into extreme and/or non-standard accretion processes near supermassive black holes. Follow-up observations with higher-resolution X-ray spectroscopy, such as with \xmm\ or \textit{XRISM}, will be essential to confirm the nature of this feature and better constrain the physical mechanisms driving it.

\end{abstract}



\section{Introduction} \label{sec:intro}
Active Galactic Nuclei (AGN) are the observational signature of supermassive black holes (SMBHs) accreting material at the centers of their host galaxies \citep[e.g.,][]{magorrian98,kormendy13,heckman14}. 
Over the past two decades, X-ray observations have revealed peculiar absorption features in AGN spectra, which are commonly interpreted as signatures of winds or outflows driven by radiation pressure on ionized gas near the accretion disc \citep[e.g.,][]{king15}.
In the most extreme cases, known as ultra-fast outflows (UFOs; \citealp[e.g.,][]{cappi09, tombesi11,tombesi13,matzeu23}), they reach velocities up to $v\sim$ 0.1--0.5$c$ \cite[e.g.,][]{chartas02,reeves03,braito18,serafinelli19,reeves20,braito22,gianolli24}.  These powerful outflows play a critical role in AGN feedback processes, as the energy and momentum they carry can impact the surrounding interstellar medium in the host galaxy. This feedback mechanism is believed to regulate both SMBH growth and star formation, making UFOs a critical phenomenon in the co-evolution of SMBHs and their host galaxies \citep[e.g.,][]{cappi06, king15, laha21, lanzuisi24}.

Another class of phenomena that can be considered analogs of UFOs are ultra-fast inflows\footnote{We adopt the definition of ``ultra-fast'' for velocities $v>0.03c$ from \citealp{tombesi11}.} (UFIs), which remain considerably less studied despite their crucial role in understanding the inner accretion dynamics near SMBHs \citep[e.g.,][]{king06,pounds18,pounds24}. 
For example, theoretical models and hydrodynamical simulations suggest that such inflows may originate from misaligned accretion discs that undergo tearing, potentially causing the disc to break and allowing material to fall inward rapidly under chaotic accretion \citep[e.g.,][]{nixon12, dogan18, pounds18}. Thus, UFIs could offer valuable insights into these non-standard accretion processes, which typically assume cold accretion flows \citep[e.g.,][]{gaspari17,kobayashi18}, posing a challenge to traditional models of circular, planar accretion discs \citep[e.g.,][]{shakura1973}.
Other possible origins for material infalling onto SMBHs include fallback from failed outflows, such as in the ``aborted jet'' or ``failed wind'' scenarios \citep[e.g.,][]{ghisellini04,progra04}, where material launched from inner winds fails to escape and returns inward. UFIs, therefore, serve as valuable probes of these mechanisms and can help advance our understanding of accretion and feedback in the innermost regions of AGN.

Only a small number of documented cases of possible UFIs exist in the literature, including, e.g., Mrk 509 \citep{dadina05},  E1821+643 \citep{yaqoob05}, Mrk 335 \citep{longinotti07}, PG1211+143 \citep{reeves05,pounds18,pounds24}, NGC 2617 \citep{giustini17}, and tentative detections from \cite{tombesi10}. In addition, these inflow detections have typically been limited to low significance, with the most evident cases reaching no more than $\sim$3--3.5$\sigma$. This is because inflows not only share variability and unpredictability with outflows, but they also appear to be rarer \citep{laha21}. This rarity limits our understanding of the phenomenon, and key aspects, such as their identification and the physics behind what drives these events, remain unclear.

Here, we present the case study of ESP~39607, a highly luminous and heavily obscured Seyfert 2 galaxy at $z=0.201$ \citep{vettolani98,landi10,ricci17}, part of the BAT AGN Spectroscopic Survey (BASS DR2; \citealp{oh18,koss22_dr2overview,koss22_dr2catalog}). In this source, we report the detection of an absorption feature in two \nustar\ observations, which we tentatively identify as a UFI.

The paper is organized as follows: in Section \S \ref{sec:data}, we describe the data used in this work; in Section \S \ref{sec:analysis}, we present our analysis; and in Section \S \ref{sec:discussion}, we discuss the main results and their implications.
Throughout this work, we assumed a $\Lambda$CDM cosmology with the fiducial parameters $H_0=70$ km s$^{-1}$ Mpc$^{-1}$, $\Omega_m=0.3$, and $\Omega_{\Lambda}=0.7$. Uncertainties are reported at a 90\% confidence level, if not stated otherwise.

\section{Data} \label{sec:data}
ESP~39607 was observed with \nustar\ for 21.2 ks in May 2023, and for 21.6 ks in August 2024 as a follow-up of the first observation.
The data were reduced using NuSTARDAS v2.1.2. Specifically, we calibrated and cleaned the downloaded datasets using the standard \texttt{nupipeline} routine.
The spectral extraction was done using the \texttt{nuproducts} command for the two FPMA and FPMB cameras.
We extracted the source spectra from a circular region with a radius of 60\arcsec. The background was extracted from two 75\arcsec\ circular regions close to the source region, on the same detector, avoiding chip gaps, and ensuring no other sources were nearby. All the spectra were binned with a minimum of 5 counts/bin, balancing information loss and possible biases in W-statistic fitting \citep{buchner23}.

For the first observation, no background flares or other issues that could significantly impact the data were identified. For the second observation, the \nustar background report\footnote{\href{https://nustarsoc.caltech.edu/NuSTAR_Public/NuSTAROperationSite/SAA_Filtering/saa_reports/}{https://nustarsoc.caltech.edu/NuSTAR\_Public/\\NuSTAROperationSite/SAA\_Filtering/saa\_reports/}} flagged the source for solar flare activity. However, the impact on the data was minimal (\nustar, private communication). We adopted a conservative approach by following the \nustar recommended guidelines for identifying and mitigating solar flare contamination\footnote{\href{https://github.com/NuSTAR/nustar-gen-utils}{https://github.com/NuSTAR/nustar-gen-utils}}. Specifically, we used the \texttt{GTI\_filter\_solar\_flare.ipynb} notebook to create custom good time intervals (GTIs) for use in the \texttt{nuproducts} routine. This resulted in the removal of less than 1\% of the total exposure from the initial extraction, confirming the minimal impact of flaring activity on the observation.

In addition to the \nustar\ data analyzed in this work, a 60.7 ks \textit{Suzaku}/XIS observation from  December 2010 is also available for ESP~39607. We used the spectra extracted by \cite{ricci17} and refer the reader to that work for details on data reduction. The same dataset is also included and shown in the broad-band spectral fitting analysis presented in Peca et al. (submitted), and we refer to both studies for additional modeling details.
Applying the same procedure described in Section~\ref{sec:analysis}, we did not find evidence of the absorption feature seen in the \nustar\ spectra within the \textit{Suzaku} data. The implications of this observation are discussed in Section~\ref{sec:discussion}.
A summary of the observations analyzed in this work is provided in Table~\ref{tab:sum_obs}.

\begin{table}[!tp]
\centering
\begin{tabular}{ccccc}
\hline
Telescope         & Obs. ID     & Date           & Exp.    & Flux   \\
\hline
\nustar           & 60860001002  & 2023-05-08     & 21.2    & 1.55$_{-0.18}^{+0.06}$   \\
\nustar           & 60161025002  & 2024-08-22     & 21.6    & 1.01$_{-0.25}^{+0.05}$    \\
\textit{Suzaku}   & 705048010    & 2010-12-19     & 60.7    & 1.89$_{-0.09}^{+0.05}$    \\
\hline
\end{tabular}
\caption{Summary of the observations used in this work. From left to right: observation ID, date of observation, exposure time in ks, and observed 2--10\,keV flux in units of $10^{-12}$\,erg\,s$^{-1}$\,cm$^{-2}$. Fluxes for the \nustar\ data are derived from the spectral analysis presented in Section~\ref{sec:analysis}, using Model 1 plus a Gaussian absorption line (see text for details). Fluxes derived with the other tested models are consistent within $\sim$1\%. The flux for the \textit{Suzaku} observation is from the best-fit results presented in Peca et al. (submitted).}
\label{tab:sum_obs}
\end{table}

\section{Spectral Analysis} \label{sec:analysis}
The spectral analysis was performed with XSPEC v12.14.0 \citep{xspec}. W-statistic \citep{cash79,wachter79} was applied. All the fits were performed in the 3--35\,keV energy range, as at higher energies the background dominates over the source emission.

\subsection{May 2023 observation}
We started by fitting the 2023 observation, which has a net (i.e., background subtracted) count rate of 0.050$\pm$0.002 $\rm ct\,s^{-1}$.
First, we modeled the source spectrum using a simple model composed of an absorbed power-law with an energy cut-off fixed at 200\,keV \citep[e.g.,][]{ricci17,zappacosta18} plus a \textit{pexrav} \citep{pexrav} component to account for reflected emission. 
For simplicity, we assumed pure reflection by fixing $R=-1$ \cite[e.g.,][]{piconcelli11,zhao19,serafinelli24}; the negative value directs the model to operate in reflection-only mode \citep[e.g.,][]{alexander11,boorman24}, while a value of one corresponds to reflection from an infinite slab illuminated by isotropic coronal emission \citep[e.g.,][]{marchesi16b,zappacosta18}.
The inclination angle was set to the default value of 60$^\circ$ \cite[e.g.,][]{ueda07,zhao20}, and the photon index $\Gamma$ was linked to that of the primary power-law component. After verifying consistency between the normalizations of the primary power-law and reflection components, we linked them to the same value to minimize degeneracies \citep[e.g.,][]{liu17,peca23}.
We refer to this configuration as Model 1.
Second, we fitted the source spectrum with the more complex, but physically motivated and self-consistent model, UXCLUMPY \citep{buchner19}. We fixed the TORsigma and CTKcover parameters to their standard values, 28 and 0.4, respectively, and the inclination angle to 60$^\circ$, as for the first model. We refer to this configuration as Model 2. 
Results from Model 2 are consistent with those from Model 1 within the uncertainties (see Table~\ref{tab:results}), and both models confirm the obscured and luminous nature of the source, with derived $N_H\simeq 3\cdot10^{23}$ cm$^{-2}$ and $L_X\simeq 5\cdot10^{44}$ erg s$^{-1}$, in agreement with those previously reported by \citet{ricci17}.

\begin{figure*}[!tp]
    \centering
    \includegraphics[width=0.494\textwidth]{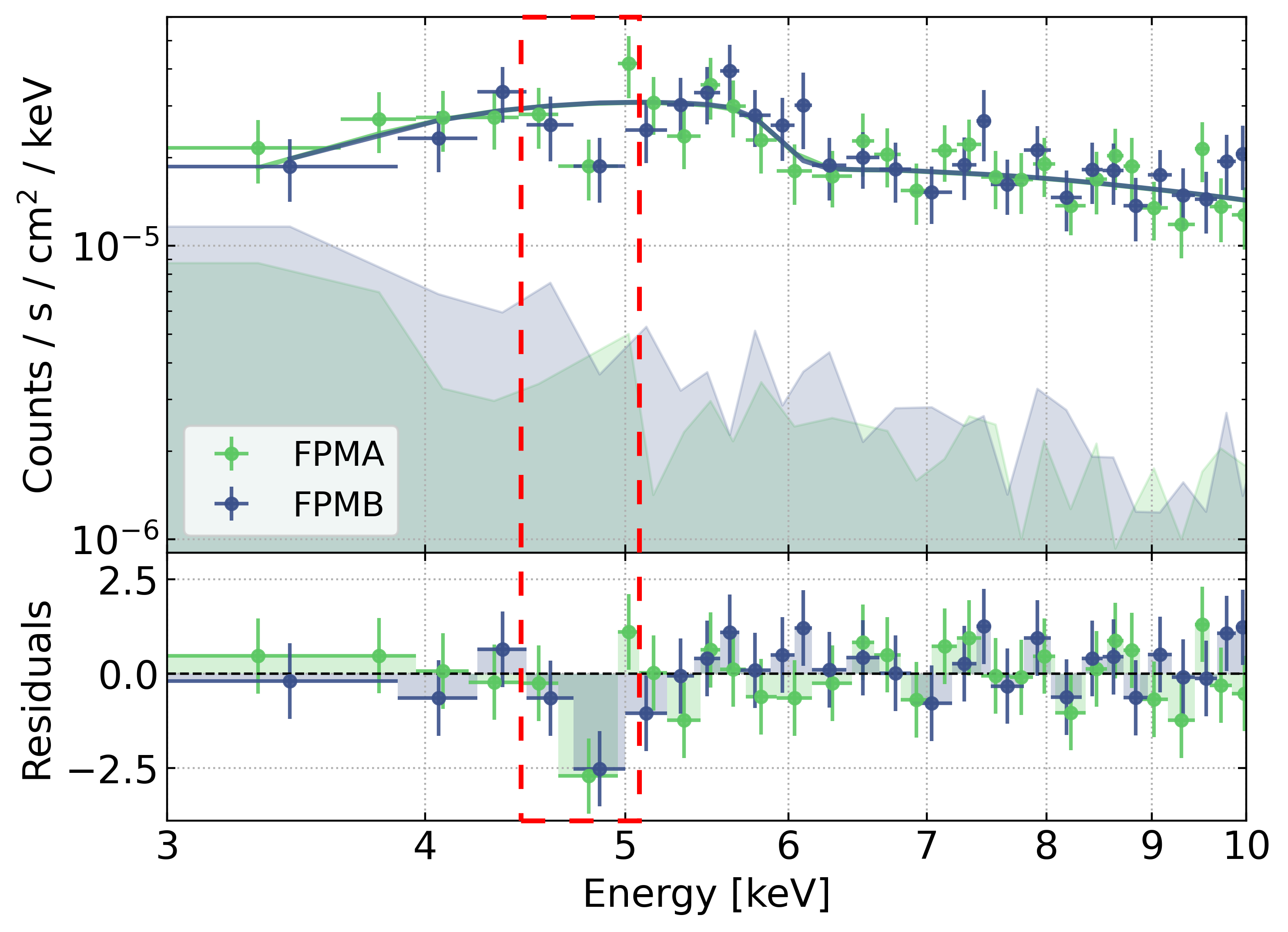}
    \includegraphics[width=0.494\textwidth]{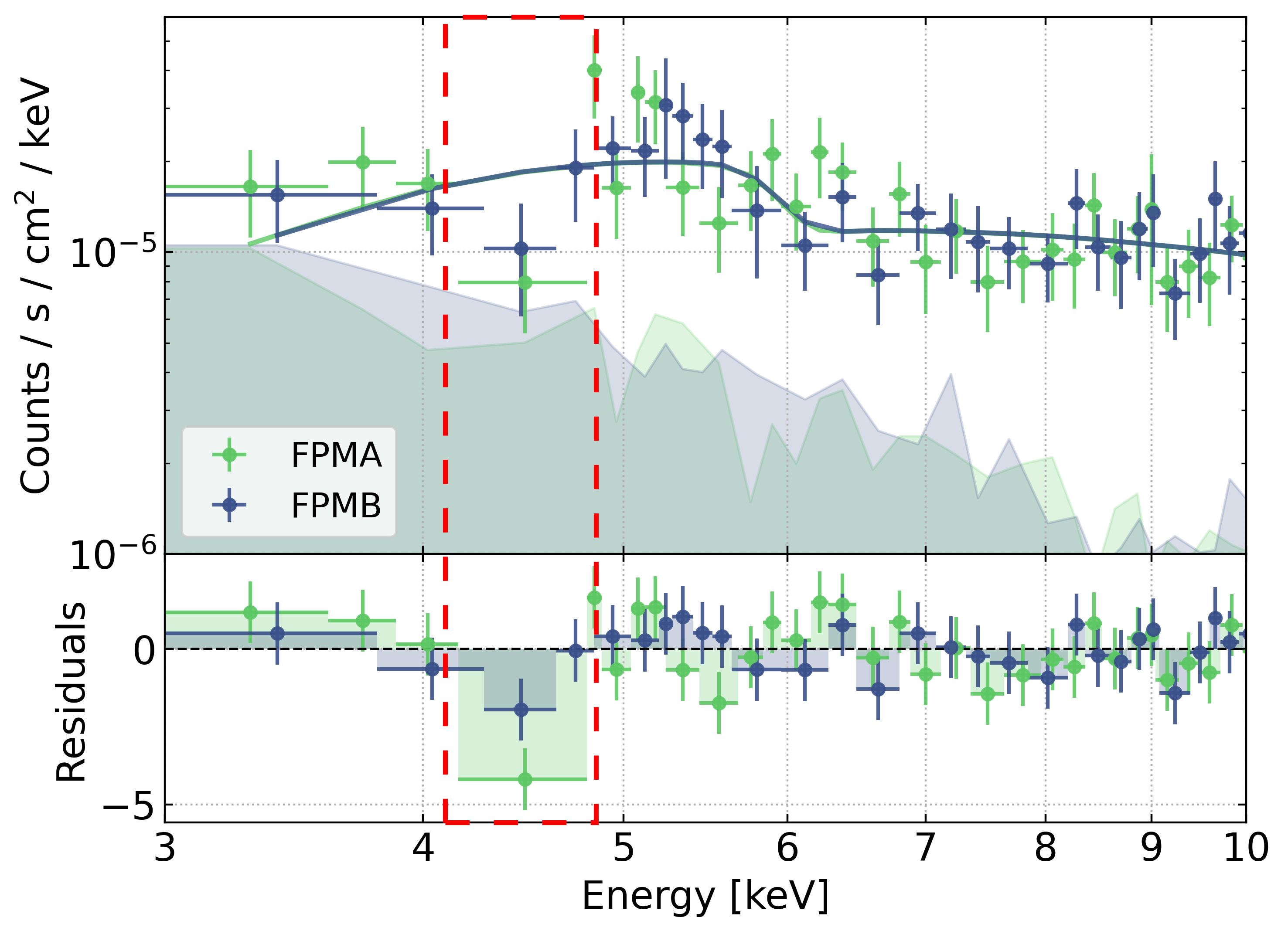}

    \caption{The best-fit results for the 2023 (left) and 2024 (right) \nustar\ spectra (3--10\,keV zoom-in) of ESP~39607, using Model 1, a simple absorbed cutoff power-law plus a reflection component. \nustar\ FPMA and FPMB cameras are shown in green and blue, respectively. The top panels show the observed data points (rebinned for graphic purposes), the best-fit model (solid lines), and the background upper range (shaded regions) for each detector, while the bottom panels show the residuals, computed as $(\mathrm{data} - \mathrm{model})/\mathrm{error}$. The red rectangles highlight the location of the absorption feature, which is not explained by the model in both observations.}
    \label{fig:12_spec}
\end{figure*}

\begin{figure}[!tp]
    \centering
    \includegraphics[width=0.47\textwidth]{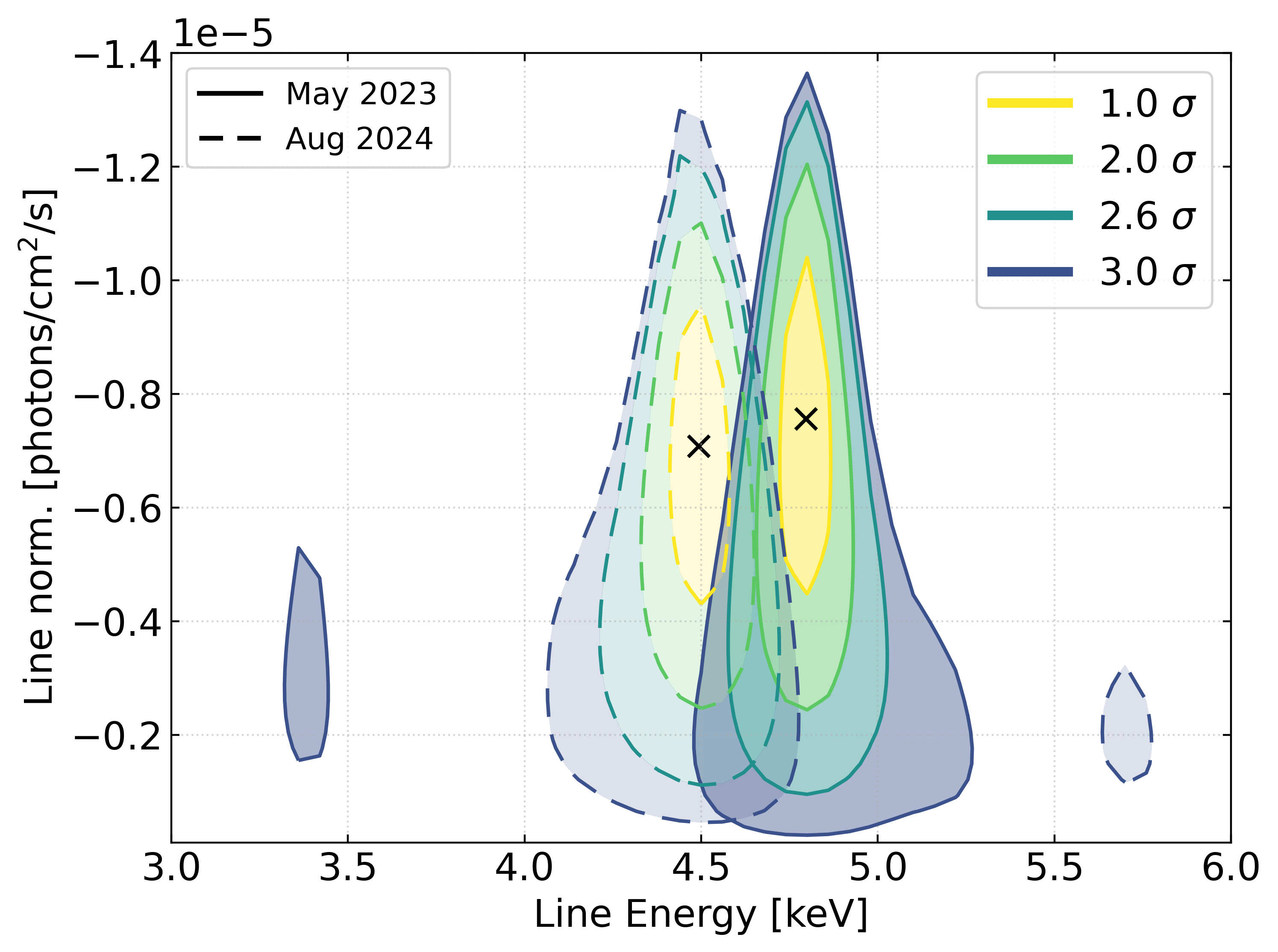}

    \caption{Combined \texttt{steppar} results for line energy and normalization of a Gaussian absorption line added to Model 1 for the 2023 (solid lines, darker colors) and 2024 (dashed lines, lighter colors) spectra (see Figure~\ref{fig:12_spec}). Contours show 68 (1$\sigma$), 95.4 (2$\sigma$), 99 ($\sim 2.6\sigma$), and 99.7\% (3$\sigma$) confidence levels. The black crosses mark the best-fit values for each observation.}
    \label{fig:12_cont}
\end{figure}

The fit residuals revealed a possible absorption feature at $\sim$4.8 keV (Figure~\ref{fig:12_spec}, left panel). To model this feature, we introduced a Gaussian line component to both models, where line normalization and energy were left free to vary. The line width was fixed to $\sigma = 10$ eV \citep[e.g.,][]{tombesi10,vignali15}, as it was found to be unconstrained at the lower end when left free, i.e., consistent with a narrow and unresolved feature. We used the XSPEC \texttt{steppar} command to map the confidence contours on line normalization and energy, and show them in Figure~\ref{fig:12_cont}. 
The inclusion of this absorption line improved the fit significantly.
For Model 1, we obtained $\Delta \text{cstat/d.o.f.} = 12.2$/2, which corresponds roughly to a fit improvement of 3$\sigma$ assuming Gaussian approximation \citep[e.g.,][]{tozzi06}. For Model 2, the improvement was $\Delta \text{cstat/d.o.f.} = 10.1$/2, i.e., $\sim$2.7$\sigma$. 
Model 1 yielded $E_{\rm line}=4.8\pm0.1$ keV and an equivalent width of $-0.21 _{-0.10}^{+0.09}$ keV, with Model 2 giving consistent results.
We summarize all the detailed results for all model configurations in Table~\ref{tab:results}.



We performed an Akaike Information Criterion test (AIC, \citealp{akaike1974}) to compare models with and without the Gaussian line. 
In brief, AIC provides a relative estimate of model quality by balancing goodness of fit against model complexity, and it is valid for both nested and non-nested models.
It is defined as $\mathrm{AIC} = -2 \ln(L) + 2k$, where $L$ is the maximum likelihood from the fit and $k$ is the number of free parameters. The model with the lowest AIC is preferred.
According to the \cite{burnham02} scale, $4 < \Delta \text{AIC} < 7$ indicates that the model without the additional Gaussian line has ``considerably less'' statistical support, while $\Delta \text{AIC} > 10$ suggests the support for the model without the line is ``essentially none''. Here, we interpret $4 < \Delta \text{AIC} < 7$ as providing ``positive'' evidence, $7 < \Delta \text{AIC} < 10$ as ``strong'' evidence, and $\Delta \text{AIC} > 10$ as ``very strong'' evidence in favor of the model with the Gaussian line.
We obtained $\Delta \text{AIC}=7.3$ for Model 1 and $\Delta \text{AIC}=5.2$ for Model 2, thereby statistically justifying the inclusion of the Gaussian line in the preferred model as strong and positive, respectively.

To further assess the reliability of the absorption feature detection, we adopted a Monte Carlo (MC) approach, as widely used in the literature to evaluate the significance of emission and absorption lines \citep[e.g.,][]{markowitz06,vignali15,costanzo22,peca23axis}. We simulated $10^5$ spectra using the best-fit models without the Gaussian line, keeping the same responses and background of the observed data. We then fitted each simulated spectrum by adding a Gaussian line and performing a scan over the 3–10 keV energy range using the \texttt{steppar} command with steps of $\Delta E=0.1$ keV. This range was chosen to avoid the drop in \nustar’s effective area at $\sim$10\,keV, which could compromise the reliability of the process \citep{peca21}. The resulting best-fit statistic is compared to the statistic from the model without the absorption line. This allows us to calculate the fraction of cases where the $\Delta \text{cstat}$ exceeds that found in the real spectrum, i.e., the fraction of cases where a fluctuation in the spectrum has a significance equal or higher than that of the absorption line in the real spectrum. This occurred in only 1.4\% of the simulations for Model 1 and 5.0\% for Model 2. In other words, the probability that the absorption feature is real is 98.6\% for Model 1 and 95.0\% for Model 2, which can be interpreted as detection significances of approximately 2.5$\sigma$ and 2.0$\sigma$, respectively.

\subsection{August 2024 observation}\label{sec:2024_obs}
Based on these results, a follow-up \nustar\ observation was conducted in August 2024.
The August 2024 observation has a much lower net count rate, 0.033$\pm$0.002 $\rm ct\,s^{-1}$. Compared to the 2023 observation, this indicates a decrease in flux of $\sim$35\% and in signal-to-noise ratio\footnote{Computed as $S / \sqrt{S+B}$, where $S$ and $B$ are the source and background counts in the source region.} from 30 to 24.
We performed the same spectral analysis as described for the 2023 observation. The spectrum and confidence contours are shown in Figure~\ref{fig:12_spec} (right panel) and Figure~\ref{fig:12_cont}, respectively.
The new observation confirmed the absorption feature, as adding a Gaussian line improved the fit by $\Delta \text{cstat/d.o.f.} = 13.1$/2 and 11.1/2 for Model 1 and Model 2, respectively, corresponding to significance levels of $\sim3.2$ and $2.9\sigma$. The AIC criterion further supports the presence of the absorption line, yielding $\Delta \text{AIC} = 9.1$ and $7.2$, indicating a ``strong'' statistical evidence in favor of its inclusion in both models.



Interestingly, compared to the 2023 observations, the best-fit line energy is shifted to a slightly lower energy of $4.5\pm0.1$ keV. However, we note that the line centroids are consistent within the 99\% confidence level, and the observed difference is also compatible with \nustar’s spectral resolution at these energies ($\sim$400 eV).
Compared to the earlier observation, the absorption feature also shows a slightly higher equivalent width, though consistent within the uncertainties. No significant variation in other model parameters was found between the two \nustar\ observations.

\begin{figure}[!tp]
    \centering
    \includegraphics[width=0.45\textwidth]{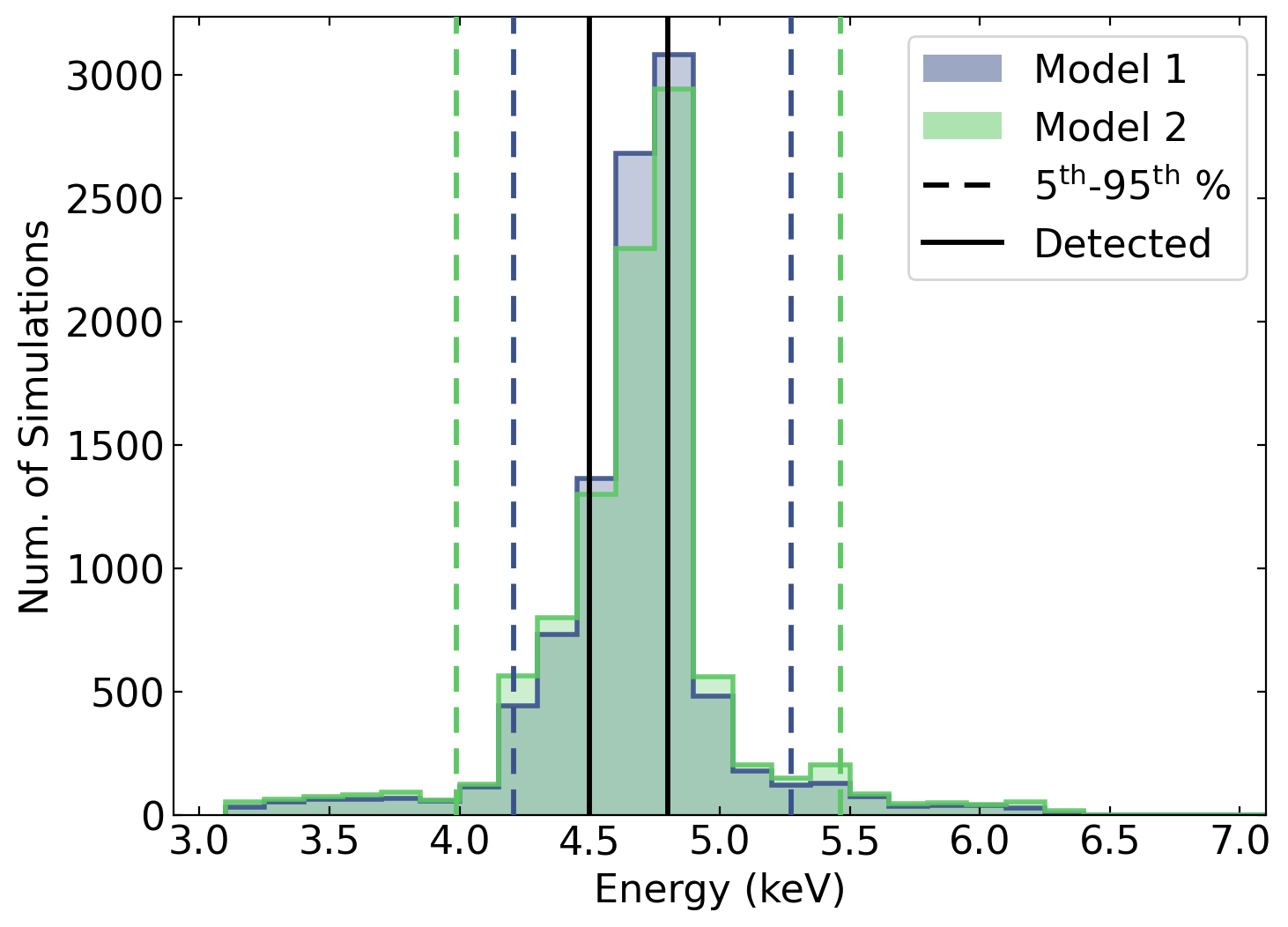} 
    \caption{Distribution of the line centroid energy obtained from $10^4$ simulated spectra based on the 2023 best-fit models, adjusted to match the lower flux and signal-to-noise ratio of the 2024 observation. Simulations were performed using Model 1 (blue) and Model 2 (green). The solid black vertical lines mark the centroids measured in the 2023 (4.8 keV) and 2024 (4.5 keV) data. Dashed colored lines indicate the 5$^{\mathrm{th}}$ and 95$^{\mathrm{th}}$ percentiles of the simulated distributions for the two models; the 16$^{\mathrm{th}}$ and 84$^{\mathrm{th}}$ percentiles are omitted for visual clarity (see text). The 2024 centroid lies well within the expected range from the simulations, supporting the interpretation that the same absorption feature is present in both observations from a statistical perspective.}
    \label{fig:centroid}
\end{figure}

We further explored the scenario where the absorption feature is the same between the two observations by performing MC-based simulations. Starting from the best-fit models of the 2023 observation, including the absorption line, we simulated $10^4$ spectra after adjusting the model normalization to match the lower flux and signal-to-noise ratio of the 2024 observation. Each simulated spectrum was then fitted using the same procedure described earlier, using the \texttt{steppar} command to scan the spectra and identify the line centroid. The resulting distribution is shown in Figure~\ref{fig:centroid}. The 4.5 keV centroid measured in the 2024 observation falls well within the 5$^{\mathrm{th}}$ percentile of the simulated distributions for both models (4.2 keV for Model 1 and 4.0 keV for Model 2), and is very close to the 16$^{\mathrm{th}}$ percentile (4.5 keV for both models), corresponding to the 90\% and 68\% confidence levels, respectively. We therefore conclude that, from a purely statistical standpoint, the absorption line in the 2024 observation is consistent with being the same as that in 2023.

\subsection{Simultaneous fitting}
We then performed a simultaneous fit of the 2023 and 2024 observations. To account for the $\sim35\%$ decrease in flux in the 2024 dataset, we introduced a multiplicative constant to allow for a different normalization between the two observations. For both models, this constant was found to be $0.68 \pm 0.04$ when including the Gaussian line, and $0.67 \pm 0.04$ when it was not.
Since the other model parameters were found to be consistent when fitting the observations independently, we linked them across the two datasets. For the absorption line, however, we allowed the energy and normalization to vary freely between the two observations. This choice was motivated by the fact that linking the line component forces the fit to settle between the two values observed in the individual fits, failing to properly capture the line profiles in either epoch. Indeed, when the line energy was linked, we obtained a centroid of $E_{\rm line} = 4.7\pm 0.1$ keV, with $\Delta \text{cstat/d.o.f.}$ values of 13.4/2 and 10.7/2 for Model 1 and Model 2, respectively, which are fit improvements comparable to those obtained when fitting the observations separately.
Instead, when allowing the line energy and normalization to vary independently, the fit significantly improved. We found $\Delta \text{cstat/d.o.f.}$ values of 25.8/4 and 21.9/4 for Model 1 and Model 2, respectively, corresponding to significances of 4.1$\sigma$ and 3.7$\sigma$. The associated $\Delta$AIC values are 17.8 and 13.9, indicating a ``very strong'' preference for the models with the Gaussian line. The best-fit values for the other model parameters remain consistent with those obtained from the individual fits and are reported in Table~\ref{tab:results}.

\begin{table*}[!ht]
\centering 
\begin{tabular}{cccccccccccc} 
\hline
Obs. & Model   & $N_{\rm H}$                  & $\Gamma$               & $L_X$               & $E_{\rm line}$ & EW                      & cstat/dof  & $\Delta$cstat & AIC    & $\Delta$AIC    & MC        \\
\hline
2023 & 1       & 34.0$_{-9.0}^{+10.9}$  & 1.95$_{-0.17}^{+0.20}$ & 4.2$_{-1.6}^{+2.9}$ & -            & -                       & 368.8/378  &              & 374.8  &            &             \\
2023 & 1+Line & 34.1$_{-9.0}^{+9.8}$   & 1.98$_{-0.17}^{+0.19}$ & 4.6$_{-1.8}^{+3.2}$ & 4.8$\pm0.1$  & -0.22$_{-0.09}^{+0.12}$ & 356.6/376  & 3.0$\sigma$  & 366.6  &  strong    & 2.5$\sigma$ \\
2023 & 2       & 26.8$_{-5.7}^{+8.9}$   & 1.82$_{-0.12}^{+0.18}$ & 6.2$_{-2.1}^{+3.9}$ & -            & -                       & 367.9/378  &              & 373.9  &            &       \\
2023 & 2+Line & 27.2$_{-6.3}^{+8.0}$   & 1.90$_{-0.16}^{+0.18}$ & 6.7$_{-2.7}^{+3.4}$ & 4.8$\pm0.1$  & -0.24$_{-0.08}^{+0.10}$ & 357.8/376  & 2.7$\sigma$  & 367.8  &  positive  & 2.0$\sigma$ \\
\hline
2024 & 1       & 37.2$_{-13.8}^{+25.6}$ & 1.94$_{-0.23}^{+0.34}$ & 2.8$_{-1.4}^{+4.3}$ & -            & -                       & 274.7/279  &              & 280.7  &            &       \\
2024 & 1+Line & 29.7$_{-10.7}^{+15.7}$ & 1.90$_{-0.20}^{+0.24}$ & 2.6$_{-1.1}^{+2.7}$ & 4.5$\pm0.1$  & -0.31$_{-0.20}^{+0.11}$ & 261.6/277  & 3.2$\sigma$  & 271.6  &   strong   & 2.4$\sigma$ \\
2024 & 2       & 31.2$_{-11.4}^{+10.1}$ & 1.81$_{-0.22}^{+0.17}$ & 4.5$_{-2.4}^{+3.7}$ & -            & -                       & 272.9/279  &              & 278.9  &            &       \\
2024 & 2+Line & 24.7$_{-8.9}^{+10.7}$  & 1.77$_{-0.20}^{+0.21}$ & 3.9$_{-1.8}^{+3.8}$ & 4.5$\pm0.1$  & -0.27$_{-0.18}^{+0.13}$ & 261.8/277  & 2.9$\sigma$  & 271.8  &   strong   & 2.2$\sigma$ \\
\hline
Both & 1       & 34.7$_{-7.7}^{+9.9}$   & 1.96$_{-0.14}^{+0.16}$ & 4.2$_{-1.4}^{+2.3}$$^{*}$ & -            & -                       & 645.0/659  &              & 653.0  &            &       \\[4pt]
\multirow{2}{*}{Both} & \multirow{2}{*}{1+Line} & \multirow{2}{*}{32.3$_{-6.9}^{+8.6}$}   & \multirow{2}{*}{1.99$_{-0.13}^{+0.15}$} & \multirow{2}{*}{4.3$_{-1.4}^{+2.2}$$^{*}$} & 4.8$\pm0.1$  & -0.20$_{-0.10}^{+0.06}$ & \multirow{2}{*}{619.2/655}  & \multirow{2}{*}{4.1$\sigma$}  & \multirow{2}{*}{635.2}  &   \multirow{2}{*}{very strong}   & \multirow{2}{*}{4.4$\sigma$} \\
     &         &                        &                        &                     & 4.5$\pm0.1$  & -0.32$_{-0.13}^{+0.12}$ &            &              &        &            &   \\[4pt]
Both & 2       & 27.4$_{-4.7}^{+7.6}$   & 1.80$_{-0.09}^{+0.17}$ & 6.1$_{-1.6}^{+4.2}$$^{*}$ & -            & -                       & 642.5/659  &              & 650.5  &            &       \\[4pt]
\multirow{2}{*}{Both} & \multirow{2}{*}{2+Line} & \multirow{2}{*}{25.9$_{-4.7}^{+6.7}$}   & \multirow{2}{*}{1.83$_{-0.10}^{+0.14}$} & \multirow{2}{*}{6.2$_{-1.6}^{+3.6}$$^{*}$} & 4.8$\pm0.1$  & -0.20$_{-0.08}^{+0.10}$ & \multirow{2}{*}{620.6/655}  & \multirow{2}{*}{3.7$\sigma$}  & \multirow{2}{*}{636.6}  &   \multirow{2}{*}{very strong}   & \multirow{2}{*}{4.2$\sigma$} \\
     &         &                        &                        &                     & 4.5$\pm0.1$  & -0.33$_{-0.10}^{+0.11}$ &            &              &        &            &   \\
\hline
\end{tabular}
\caption{Results of the spectral analysis for the two observations and the two models tested. Columns show the observation date (where ``Both'' refers to the simultaneous fit of the two epochs); the model applied for the spectral fitting where ``1'' is for the simple absorbed power-law plus reflection, ``2'' for UXCLUMPY, and ``+Line'' indicates the results when including the Gaussian line; the absorption ($N_{\rm H}$) in units of $10^{22}$ cm$^{-2}$; photon index ($\Gamma$); 2--10\,keV, absorption-corrected luminosity ($L_{\rm X}$) in units of 10$^{44}$ $\rm erg\,s^{-1}$; line centroid ($E_{\rm line}$) in keV; line equivalent width (EW) in keV; cstat value over degrees of freedom for the best-fit model, significance of the Gaussian line according to the $\Delta$cstat value (under Gaussian approximation) between the model with and without the line; AIC value; $\Delta$AIC score when adding the Gaussian line; and significance of the absorption line based on our MC-based simulations. The observed fluxes are reported in Table \ref{tab:sum_obs}.
In the simultaneous fits, luminosities marked with an asterisk ($^{*}$) refer to the first observation only; the luminosity of the second observation should be scaled down by a factor $0.68\, (0.67) \pm 0.04$ for the models with (without) the line, as discussed in the text. }
\label{tab:results}
\end{table*}

To further quantify the significance of detecting the same absorption feature in both observations, we performed MC-based simulations following the same procedure used for the individual epochs. Starting from the best-fit continuum models (i.e., without any absorption line), we simulated $2 \cdot 10^5$ spectra for each observation. For each simulated pair, we searched for spurious absorption features in each spectrum with a $\Delta \text{cstat}$ equal to or greater than that observed when including the Gaussian line. We then required that these features appear in both spectra within an energy separation of 0.3 keV, corresponding to the difference between the observed centroids. 
The resulting false positive probabilities translate into detection significances of 4.4$\sigma$ for Model 1 and 4.2$\sigma$ for Model 2, reinforcing the robustness of the detection across both epochs.

\section{Discussion} \label{sec:discussion}

\subsection{A possible ultra-fast inflow?}
We first explore the interpretation that the observed absorption feature arises from an ionized inflow, a plausible physical scenario for an absorption line at these energies \citep[e.g.,][]{longinotti07,tombesi10,pounds18,pounds24}. At the source redshift ($z = 0.201$), the observed 4.8 keV feature corresponds to $\sim$5.8 keV in the rest frame. Given the observed energy range, the most likely candidates are \fexxv\ (rest-frame, 6.70 keV) or \fexxvi\ (6.97 keV) K$\alpha$ transitions, with iron favored due to its high cosmic abundance compared to other elements and frequent detection in X-ray spectra \citep[e.g.,][]{reeves05,longinotti07}. We focus exclusively on the K$\alpha$ transition, as a reliable identification of K$\beta$ would require simultaneous detection of both transitions, which we do not observe. 
For the following analysis, we rely on the 2023 observation, which provides the highest signal-to-noise ratio. 

\subsubsection{Inflow velocity and position}

We estimated the inflow velocity using the the relation $(1+z_0)=(1+z_a)(1+z_c)$ \citep[e.g.,][]{tombesi11,serafinelli23,bertola20}, where $z_0$ is the observed redshift of the absorber, $z_a$ is the intrinsic absorber redshift, and $ z_c $ is the cosmological redshift of the source. The inflow velocity $v$ is then derived from $(1+z_a)=\sqrt{\frac{1-\beta}{1+\beta}}$, where $\beta = v/c$. This calculation indicates an ultra-fast inflow (UFI) with a velocity of $\sim$0.20$c$ when assuming \fexxvi\ K$\alpha$ and $\sim$0.15$c$ assuming \fexxv\ K$\alpha$, consistent with previous UFI studies \citep[e.g.,][]{dadina05,giustini17,pounds18}. 
We note that, given the limited spectral resolution of our data ($\sim$400 eV), we cannot distinguish between these two lines, and therefore the detected feature could be a blend of both transitions.

To test the physical consistency of these findings, we compared the inferred inflow velocity with a simple model describing radially infalling gas under gravitational attraction, influenced by radiation pressure, following \cite{longinotti07}. In this scenario, we calculate the minimum radius at which the observed inflow velocity equals the gravitational free-fall velocity. This radius can be expressed as
\begin{equation}
    R = \frac{2\, G \, M_{\mathrm{BH}}}{v^2} \cdot P_{\rm rad} ,
\end{equation}
where $G$ is the gravitational constant, $v$ is the inflow velocity, $M_{\mathrm{BH}}$ is the black hole mass, and $P_{rad}=(1-L_{\rm bol}/L_{\rm Edd})$ is the radiation pressure. $L_{bol}$ is derived from the average X-ray luminosity between the two used models, using the bolometric correction of \cite{duras20}.
For \fexxvi\ K$\alpha$ (\fexxv\ K$\alpha$), this yields $R=22\,\mathrm{(34)} R_g$, where $R_g=G\cdot M_{\mathrm{BH}}/c^2$ is the gravitational radius. When radiation pressure is neglected, the corresponding values are $R=55\,\mathrm{(89)} R_g$. These distances are consistent with those found in the literature for UFIs \citep[e.g.,][]{longinotti07,pounds18}.
Assuming a black hole mass of $\log M_{\mathrm{BH}}/{M_{\odot}} = 8.26$, derived from the velocity dispersion method described in \citet{koss22_veldisp} and included in the forthcoming BASS DR3 (Koss et al., in prep.), the corresponding physical distances are $R=2\,\mathrm{(3)} \cdot 10^{-4}$ pc and $R=5\,\mathrm{(8)} \cdot 10^{-4}$ pc for the two cases, respectively.

\subsubsection{Photoionization modelling}
We then employed a more physically motivated approach by modeling the absorption feature using the XSTAR v2.59 photoionization code \citep{xstar1,xstar2}. Following \cite{tombesi11}, we assumed a power-law with $\Gamma = 2$, gas temperature T=10$^6$K \citep{nicastro99,bianchi05}, and a gas density of $n=10^{10}$ cm$^{-3}$. Solar abundances were adopted from \cite{asplund09}. We simulated three grids\footnote{We used the MPI C++ parallelization tool provided by \cite{mpi_xstar} to run XSTAR.} with different turbulence velocities, $v_{turb}=$ 100, 1000 and 5000 $\rm km\,s^{-1}$, and a fixed covering factor of 1 \citep[e.g.,][]{pounds18}. The additional free parameters given by these grids are the column density of the absorbing material ($N_{\rm H,abs}$), the redshift of the absorber ($z_0$), and the ionization parameter ($\xi$). The latter is defined by $\xi = \frac{L_{\rm ion}}{n r^2}$ (erg s$^{-1}$ cm), where $L_{\rm ion}$ is the ionizing luminosity of the source integrated between 1 and 1000 Ryd (1 Ryd = 13.6 eV), $n$ is the gas density, and $r$ is the distance between the absorber and the central source \citep{tarter69}.

\begin{figure}[!tp]
    \centering
    \includegraphics[width=0.47\textwidth]{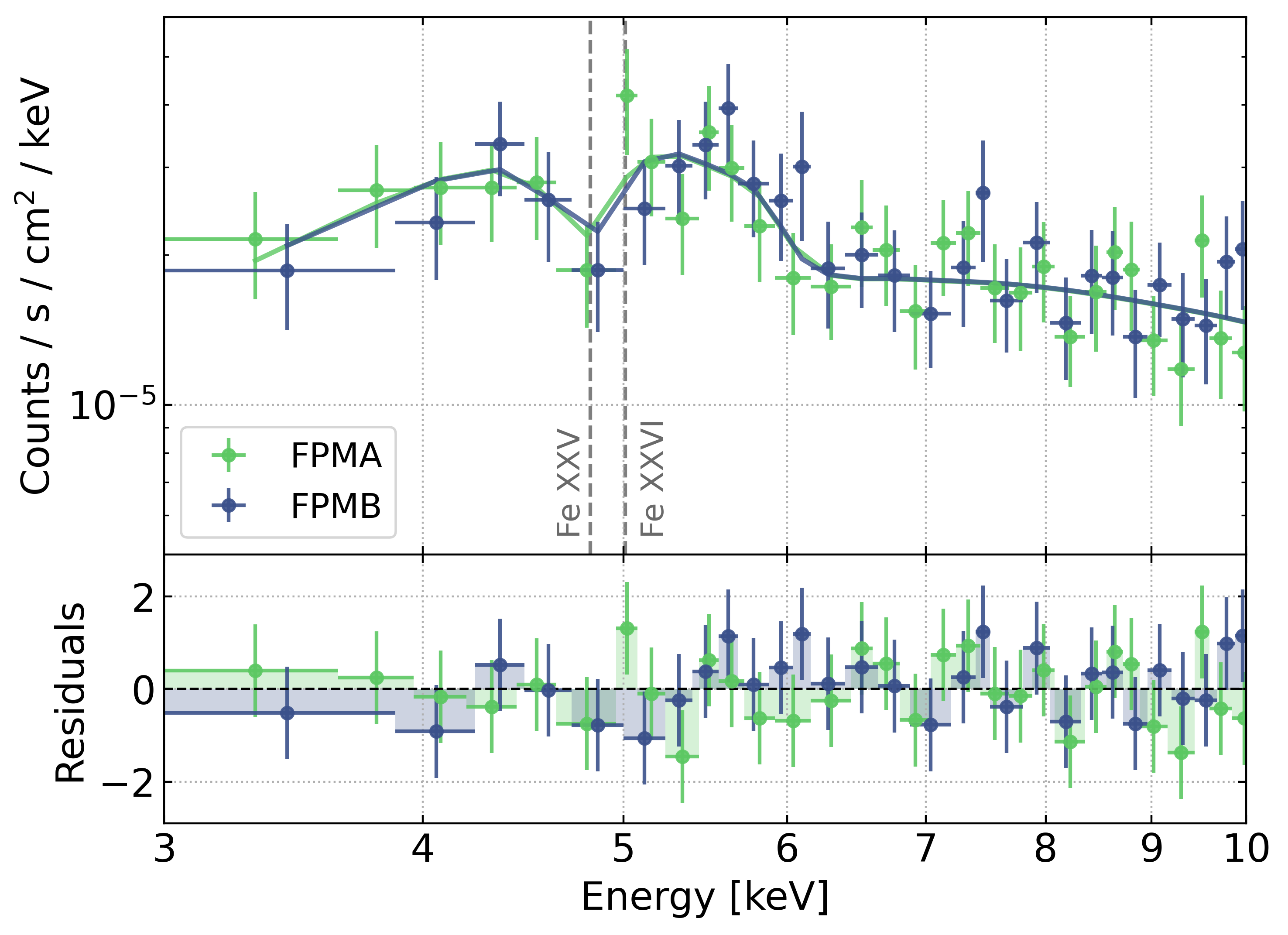}

    \caption{3--10\,keV zoom-in spectrum of the 2023 observation, showing out best-fit for Model 1 combined with the XSTAR grid with $v_{turb}$=5000 $\rm km\,s^{-1}$ (line styles and colors as in Figure~\ref{fig:12_spec}).  
    Two grey dashed vertical lines indicate the observed-frame positions of the \fexxv\ and \fexxvi\ K$\alpha$ transitions, which are blended, with the dominant contribution from \fexxv. Note that this spectrum is rebinned as in Figure~\ref{fig:12_spec} (left panel), but the two lines remain blended even in the unbinned spectrum.}
    \label{fig:xstar_results}
\end{figure}

We obtained the best results using the XSTAR grid with a turbulence velocity of 5000 $\rm km\,s^{-1}$, which provided both constrained parameters and the most improvement to the fit. The results are summarized in Table~\ref{tab:results_xstar}, and our best-fit result for Model 1 is shown in Figure~\ref{fig:xstar_results}. 
For both models we obtained consistent results, with an ionization parameter $\log \xi \simeq 3$ and an absorbing column density for the inflowing material of $N_{\rm H,abs} \simeq 2 \times 10^{23}$ cm$^{-2}$, with a redshift of $z_0 \simeq 0.39$, corresponding to an inflow velocity of  $\sim$0.15$c$. According to the XSTAR modeling, the absorption feature at this redshift is primarily produced by the \fexxv\ K$\alpha$ transition. As shown in Figure~\ref{fig:xstar_results}, the observed absorption feature is reproduced by the model as a blend of \fexxv\ and \fexxvi\ K$\alpha$ lines, with \fexxv\ K$\alpha$ providing the dominant contribution. The inclusion of the XSTAR component improves the fit by $\Delta$cstat/d.o.f. = 7.8/3 and 8.3/3 for Models 1 and 2, respectively, corresponding to $\sim 2\sigma$ confidence for three additional free parameters.
The slightly lower significance, compared to the models with the Gaussian line in absorption, arises because the XSTAR grid introduces three free parameters ($N_{\rm H,abs}$, $z_0$, and $\xi$) versus only two for the Gaussian line (line energy and normalization). Additionally, XSTAR accounts for all resonance absorption lines and edges from elements up to $Z=30$ in multiple ionization states, among which only the feature presented in this work is statistically significant in our \nustar\ dataset.

For comparison, the 1000 $\rm km\,s^{-1}$ grid produced results broadly consistent with the 5000 $\rm km\,s^{-1}$ case, but with larger uncertainties and poorer fits ($\Delta$cstat/d.o.f. = 5.0/3 and 6.3/3 for Models 1 and 2, respectively), as it failed to model the full depth of the absorption feature. For the same reason, the 100 $\rm km\,s^{-1}$ grid did not allow us to constrain any of the absorber parameters.

\begin{table*}[!ht]
\centering\hspace{-2cm}
\begin{tabular}{cccccccccc}
\hline
Obs. & Model     & $N_{\rm H}$          & $\Gamma$               & $L_X$                & $N_{\rm H,abs}$   & $\log \xi$            & $z_0$                   & cstat/dof  & $\Delta$cstat\\
\hline
2023 & 1+XSTAR & 35.7$_{-4.7}^{+6.7}$   & 1.99$_{-0.07}^{+0.05}$ & 5.0$_{-2.9}^{+1.5}$  & 19$_{-12}^{+27}$  & 3.0$_{-0.2}^{+0.5}$   & 0.39$_{-0.03}^{+0.04}$  & 361.0/375  & 2.0$\sigma$\\
2023 & 2+XSTAR & 27.6$_{-4.5}^{+9.5}$   & 1.96$_{-0.10}^{+0.15}$ & 7.5$_{-3.2}^{+2.1}$  & 19$_{-14}^{+39}$  & 3.1$_{-0.9}^{+0.6}$   & 0.39$_{-0.04}^{+0.04}$  & 359.6/375  & 2.1$\sigma$\\
\hline
\end{tabular}
\caption{Results of the spectral analysis using the grid with $v_{turb}=5000$ $\rm km\,s^{-1}$ produced with the XSTAR code. From left: observation date; model applied for the spectral fitting where ``1'' is for the simple absorbed power-law plus reflection, and ``2'' for UXCLUMPY; absorption ($N_{\rm H}$) in units of $10^{22}$ cm$^{-2}$; photon index ($\Gamma$); 2--10\,keV, absorption corrected luminosity ($L_{\rm X}$) in units of 10$^{44}$ $\rm erg\,s^{-1}$; absorbing column density of the material ($N_{H,abs}$) producing the absorption line, in units of $10^{22}$ cm$^{-2}$; logarithm of the ionization parameter ($\xi$) of the absorbing gas; redshift of the absorber ($z_0$); cstat statistic value over degrees of freedom, and the improvement to the models when adding the XSTAR grid using the $\Delta$cstat criterion.}
\label{tab:results_xstar}
\end{table*}

\subsubsection{On the Suzaku 2010 observation}
As described in Section~\ref{sec:data}, an archival \textit{Suzaku} observation from 2010 is available for ESP~39607. These data were initially reduced by \citet{ricci17} as part of a broad AGN characterization study, and have been included in a multi-epoch AGN analysis by Peca et al. (submitted), which also incorporates the \nustar\ observations presented here. In particular, the Peca et al. analysis confirms the spectral results reported in this work and classifies the source as variable in flux, with the 2010 \textit{Suzaku} observation showing a 2--10\,keV flux $\sim$20\% higher than that of the 2023 \nustar\ observation.
Peca et al. also show that soft X-ray components, such as thermal emission and scattering from the primary power-law emission, are too faint to affect the hard X-ray spectra analyzed here, with negligible impact above $\sim$3\,keV in the observed frame. We refer the reader to this work for further details.

Applying the same analysis and fitting procedure outlined in Section~\ref{sec:analysis}, we found no significant evidence of the absorption feature in the \textit{Suzaku} data. The most plausible explanation for this absence is the intrinsic variability of the infalling material. Indeed, observational studies have shown that inflows and outflows commonly vary on timescales ranging from days to years \citep[e.g.,][]{tombesi10,pounds18,laha21,braito22}, which is not at odds with the lack of detection in the 2010 observation and supports the interpretation of the feature as a transient episode of UFI.

\subsubsection{The physical mechanism that drives the inflow}
Several physical mechanisms could explain the origin of the transient UFI detected in this source. One possibility is that the inflowing material results from instabilities in a misaligned accretion disk. Theoretical models and hydrodynamical simulations suggest that such disks can undergo tearing and fragmentation, causing portions of the disk to break apart into rings or clumps that decouple from the main flow \citep[e.g.,][]{nixon12,dogan18}. These fragments can then fall inward rapidly on plunging or eccentric trajectories, leading to short-lived inflow events. This form of chaotic accretion deviates from the standard picture of smooth, circular disk accretion \citep[e.g.,][]{shakura1973}, and is typically associated with cold, transient events \citep[e.g.,][]{gaspari17,kobayashi18}. 

Alternatively, the inflow could result from material that was initially launched outward but failed to escape the gravitational pull of the central SMBH \citep[e.g.,][]{giustini17}. In the ``aborted jet'' scenario \citep{ghisellini04}, compact blobs of plasma are intermittently ejected from the inner accretion flow but lack the energy to fully escape. These blobs may rise to a certain distance before falling back, potentially interacting with new ejecta and producing shocks contributing to the X-ray emission. A related possibility is the ``failed wind'' scenario \citep{progra04}, in which the absorbing material is part of an ongoing wind that becomes over-ionized by the intense continuum radiation. As the wind loses momentum, it stalls and eventually falls back toward the disk plane. In both cases, the inflow is expected to be highly variable and short-lived, with characteristic timescales on the order of the dynamical or free-fall time. 

Under these scenarios, the emergence of the absorption feature in the 2023/2024 \nustar\ data, following its absence in the 2010 \textit{Suzaku} observation, is consistent with the expected episodic nature of such events, which are predicted to vary on timescales of years or less. Additionally, the accompanying flux variations of $\sim$20\% between the 2010 \textit{Suzaku} and 2023 \textit{NuSTAR} observations, and $\sim$35\% between the 2023 and 2024 \textit{NuSTAR} observations, further support these interpretations, as such changes are likely to be expected for disk instabilities or fallback processes \citep[e.g.,][]{ghisellini04,nixon12}.

\subsection{Other scenarios}

While our primary interpretation of the observed feature is an ionized iron absorption line from inflowing material, a scenario supported by consistency with previous UFI studies \citep[e.g.,][]{dadina05,reeves05,giustini17,pounds18}, in this Section we consider alternative potential explanations.

One possibility is that the absorption originates from the Warm-Hot Intergalactic Medium (WHIM). However, its variability between the two \nustar\ observations and absence in the \textit{Suzaku} data is inconsistent with the typically stable nature of WHIM absorption \citep[][]{nicastro18,nicastro23}. Moreover, the column densities required to reproduce the observed feature ($N_{H,abs}\simeq 2 \times 10^{23}$ cm$^{-2}$) are much larger than those typically associated with diffuse local gas ($\lesssim 10^{21}$ cm$^{-2}$; \citealp[e.g.,][]{zappacosta10,nicastro18,nicastro23}). 
An absorption line in the 4--5\,keV range in the spectrum of an AGN at $z = 0.201$ is also unlikely to originate from the WHIM. Indeed, the most abundant WHIM tracers in absorption, such as O\,{\sc vii}, O\,{\sc viii}, and Ne\,{\sc ix} \citep[e.g.,][]{kaastra06, nicastro18,nicastro23}, produce absorption features at softer X-ray energies ($\lesssim1$ keV). 
These considerations make it unlikely that the observed feature is related to the WHIM.

Another potential explanation is that the feature arises from an {\it outflow} involving elements lighter than iron \citep[e.g.,][]{gupta15,reeves20,lanzuisi24}. While possible in principle, most lighter elements produce K$\alpha$ transitions at significantly lower energies, such as Ca\,{\sc xx} (rest-frame, 4.11 keV), Si\,{\sc xiv} (2.00 keV), S\,{\sc xvi} (2.62 keV), and Mg\,{\sc xii} (1.47 keV). To explain an absorption line observed at 4.8 keV, the required outflow velocities for S\,{\sc xvi}, and Mg\,{\sc xii} ions would need to exceed 0.75$c$, making this scenario highly unlikely. Si\,{\sc xiv} would require an outflow velocity of $\sim$0.65$c$, a velocity that has only been observed in a handful of exceptional sources and exclusively associated with highly ionized iron transitions \citep[e.g.,][]{chartas09,lanzuisi12,luminari23}. The only ion among these with a rest-frame energy close enough to match the observed line with a more moderate outflow velocity is Ca\,{\sc xx}, which would require a velocity of $\sim$0.33$c$. However, calcium is $\sim$15 times less abundant than iron \citep{asplund09}, making a strong, isolated Ca\,{\sc xx} absorption feature difficult to justify. 
To further test the possibility of outflowing elements lighter than iron, we performed a blind velocity search using the XSTAR model, allowing the redshift parameter to vary into the negative velocity regime (i.e., outflows). We tested both the 5000 and 1000 $\rm km\,s^{-1}$ turbulence grids, using the \texttt{steppar} command with a step of $\Delta z = 0.01$. No significant improvement in fit statistics was found, thus providing no support for this scenario.

Based on these analyses, we conclude that the UFI interpretation is the most plausible explanation for the observed absorption feature, although we cannot entirely rule out alternative scenarios with the current data.



\section{Summary and conclusions}

We presented the detection and analysis of an absorption feature in the X-ray spectra of ESP~39607, a Seyfert 2 galaxy at $z=0.201$, observed with \nustar\ in May 2023 and August 2024. Based on our best-fit modeling, we associate this feature with an ultra-fast inflow (UFI) at a velocity of $\sim$0.15$c$, primarily produced by the \fexxv\ K$\alpha$ transition, with a blended contribution from \fexxvi\ K$\alpha$. The observed 4.8 keV feature corresponds to $\sim$5.8 keV in the rest frame of the source.
Below, we summarize our findings:

\begin{itemize}
    
    \item The absorption feature was consistently detected at 4.8 keV in both 2023 and 2024 \nustar\ observations. Although the second observation yielded a slightly different best-fit energy (4.5\,keV), simulations demonstrate that this shift is not statistically significant, reinforcing the consistency of the detections. Individually, each detection has a significance that ranges between 2 and 3$\sigma$, as estimated through MC simulations and other statistical tests. When the two observations are considered together, the significance exceeds 4$\sigma$.
    
    \item The absorption feature was first fitted with a Gaussian absorption line, combined with two different continuum models that describe the rest of the spectrum. Both models yielded consistent results. Based on the best fits, the line is tentatively interpreted as a redshifted \fexxv\ K$\alpha$ or \fexxvi\ K$\alpha$ transition, corresponding to an inflow velocity of $\sim$0.15–0.20$c$.
    
    \item A simple spherical model for radially infalling gas suggests that the inflow originates at 22 (34) $R_g$ for \fexxvi\ (\fexxv), or at 55 (89) $R_g$ if radiation pressure is considered. These estimates are consistent with values reported in the literature.
    
    \item When fitting the data with XSTAR photoionization models, the best-fit was obtained using a grid with a turbulence velocity of 5000 $\rm km\,s^{-1}$, where the absorption feature is dominated by a strong \fexxv\ K$\alpha$ line, blended with a minor contribution from \fexxvi K$\alpha$. The derived parameters, $\log \xi \sim 3$, $N_{\rm H,abs} \sim 2 \times 10^{23}$ cm$^{-2}$, and an inflow velocity of $\sim$0.15$c$, are overall in agreement with previous UFIs studies.
    
    \item We tested alternative interpretations to explain the origin of the absorption feature, including the presence of WHIM signatures and outflows from lighter elements. Based on these analyses, we conclude that the ultra-fast inflow interpretation is the most plausible explanation.
    
\end{itemize}

Although our results favor an inflow interpretation, we acknowledge the limitations of the current data. While \nustar\ enabled the detection of the absorption feature, its $\sim$400 eV energy resolution prevents distinguishing between closely spaced lines such as \fexxv\ and \fexxvi\ K$\alpha$ transitions, and limits our ability to characterize the absorber in greater detail. 
To address this limitation, we have secured observing time with \xmm\ (AO 24, PI A. Peca). This \xmm\ observation is expected to provide significantly improved sensitivity and spectral resolution ($\sim100$ eV), making it ideally suited for confirming the absorption feature and better constraining the blended \fexxv\ and \fexxvi.
Furthermore, the newly launched \textit{XRISM} mission \citep{xrism}, with its unprecedented energy resolution ($\sim$5 eV), could enable detailed characterization of the absorber. This would be especially valuable if the intrinsic turbulent velocity is lower than measured here, potentially allowing \textit{XRISM} to resolve the individual \fexxv\ and \fexxvi\ components. Together, these missions will offer complementary insights into the nature and variability of this feature, thereby improving our broader understanding of the role that UFIs play in AGN accretion and feedback processes.

\section*{acknowledgments}
We acknowledge the anonymous referee for the valuable comments that improved the quality of the paper. A.P. acknowledges D. Costanzo for the useful discussions. A.P. and M.K. thank the \nustar\ team for the helpful discussions on the observations and the data reduction of ESP~39607.  A.P. and M.K. acknowledge support from NASA through ADAP award 80NSSC22K1126 and NuSTAR grants 80NSSC22K1933 and 80NSSC22K1934. R.S. acknowledges funding from the CAS-ANID grant number CAS220016. C.R. acknowledges support from Fondecyt Regular grant 1230345, ANID BASAL project FB210003, and the China-Chile joint research fund. \vspace{1mm}


\facilities{\nustar, \textit{Suzaku}}

\software{
          NuSTARDAS v2.1.4\footnote{https://heasarc.gsfc.nasa.gov/docs/nustar/analysis/},
          XSPEC v12.14.0 \citep{xspec},
          XSTAR v2.59 \citep{xstar1,xstar2},
          Astropy v7.0.1 \citep{astropy:2013,astropy:2018,astropy:2022},
          Matplotlib v3.10.1 \citep{matplotlib},
          Pandas v2.1.4 \citep{pandas1,pandas2}.
          }

\bibliography{sample631}{}
\bibliographystyle{aasjournal}



\end{document}